*5Gperf: signal processing performance for 5G*
*Improved algorithms and SIMD software acceleration for base-stations*

Gaétan HAINS, Wijnand SUIJLEN, LIANG Wenliang and WU Zixu

# Technical Report PADAL-TR-2018-2

2018-2-7

Huawei Technologies
2012Labs/CSI/DPSL/PADAL

Huawei Paris R&D Center

# 5Gperf: signal processing performance for 5G

## improved algorithms and SIMD software acceleration for base-stations

Gaétan Hains and Wijnand Suijlen
Huawei Parallel and Distributed Algorithms Lab.
Paris Research Center, Boulogne-Billancourt, France

LIANG Wenliang and WU Zixu
Huawei 5G Research Department
Shanghai, P.R.C.

*Abstract*—The 5Gperf project was conducted by Huawei research teams in 2016-17. It was concerned with the acceleration of signal-processing algorithms for a 5G base-station prototype. It improved on already optimized SIMD-parallel CPU algorithms and designed a new software tool for higher programmer productivity when converting MATLAB code to optimized C.

## I. Introduction

As a leading vendor of wireless telecommunication systems, Huawei/CRI (Central Research Institute)'s Wireless Technology Lab is developing a 5G base-station prototype and has demonstrated its very high performance based on MIMO technology [1,2,3]. Base-station power consumption and throughput critically depends on the efficiency of the signal-processing system. Its algorithms are designed by wireless signal experts, usually in MATLAB and then have to be converted to high-performance sequential C, a labor-intensive process of up to one man*month per new algorithm or pipeline module version. The 5Gperf project has been a collaboration with Huawei's CSI (Central Software Institute)'s Paris team for improving key algorithms and designing a software tool to improve human productivity in high-performance C codes.

This paper summarizes the project and its results.

## II. 5G base station prototype

A new 5G base-station prototyped, built and tested. Its signal processing system is a pair of two algorithm pipelines for processing signal packets as shown in figure 1. Each stage in the pipeline implements a specialized algorithm and the system throughput is limited by the speed of each one. It currently runs on a hardware configuration of five Huawei E9000 blade servers connected by Infiniband. One pipeline instance per CPU core is running in continuous mode.

Each algorithm has been carefully designed in MATLAB to maximize signal quality. It has then been converted to optimized sequential C code to become the compute kernel that implements the corresponding pipeline stage. This process is too labor-intensive and the results not always optimal because of the complex interplay between hardware architecture and relatively-small compute kernels. The 5Gperf project has scrutinized some performance-critical algorithms and provided a new software tool for improving the MATLAB-to-C conversion's productivity and code performance.

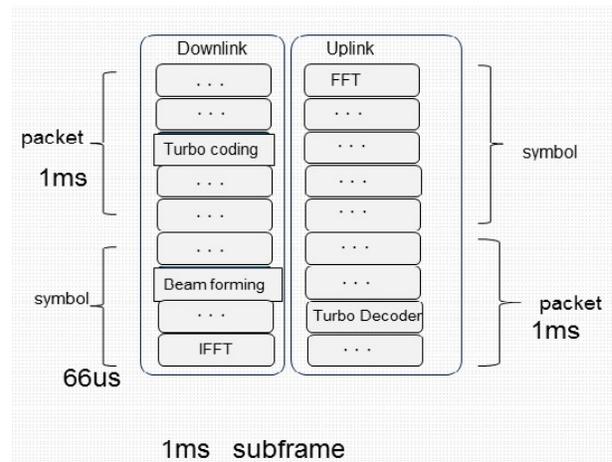

Fig. 1. The base-station's signal-processing pipeline

Figure 2 illustrates one of the prototype field-tests [9]. The top-right image shows 26 cell-phone stands like the one in the bottom-right image. They were all connected to the antenna array shown in the top-left image. The antenna array is less than 1m wide. It features 4 x 8 x 2 = 64 transceivers operating over a radio frequency band of 100MHz.

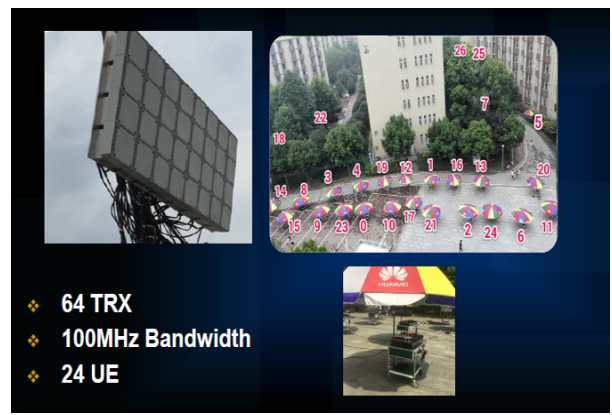

Fig. 2. The 5G base-station prototype

## III. Signal Processing, New SIMD-Optimized Algorithms

Some of the most intensive algorithms in the signal-processing pipeline have been analyzed to improve their performance on the current Intel hardware. They have been either hand-optimized or replaced by a specific Intel library if appropriate. It was decided early on that multicore parallelism could not be used effectively within each algorithm: the matrices being handled are too small to amortize RAM access times. So all optimizations had to come from reducing the total number of sequential steps and/or optimizing SIMD instruction parallelism.

Further algorithms have been re-implemented with portable algorithm building blocks provided by the new software tool described in the next section (Figure 3).

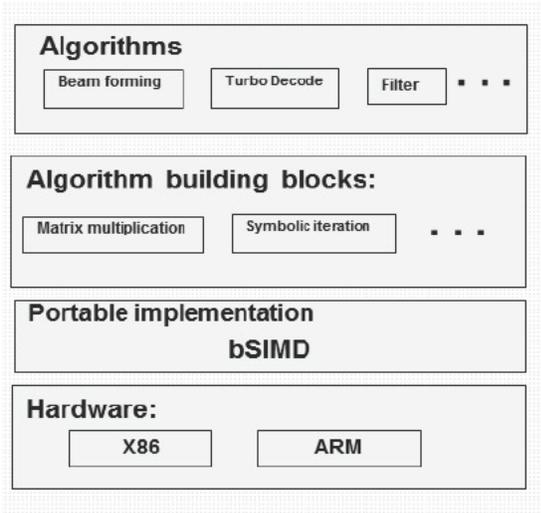

Fig. 3. The 5Gperf project stack

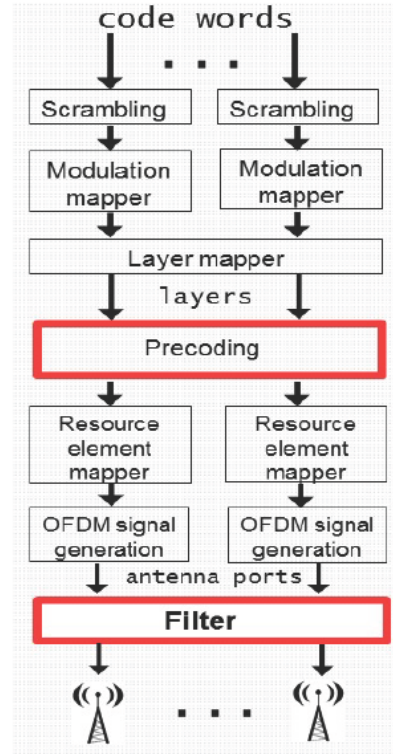

Fig. 4. Filter operations

The filter algorithm computes

$$\text{IFFT}(\text{MULT}(\text{FFT}(s)))$$

where MULT is the pointwise multiplication by the FFT of the filter sequence and FFT sizes are all 2048. Figure 4 illustrates the filter processing pipeline (redrawn from textbook [7]).

Its speed hinges on that of the FFT and we obtained good accelerations by replacing a hand-coded FFT by Intel's FFT library. The speed improvement is due to Intel's optimization for the Haswell architecture. An average speedup of 1.4 x has been recorded:

| Original (µs) | New (µs) | Speedup factor |
|---|---|---|
| 27 | 19 | 1.4 |
| 27 | 19 | 1.4 |
| 27 | 18 | 1.5 |

The beam forming algorithm computes many products $W \times S$ where W the precoding matrix is constant and S the symbols matrix is changing. A row $i$ of W corresponds to one antenna (here there are 64). Matrix S has $f$ rows, one per parallel flow, and b columns, one per resource element. By definition of matrix product each of W's rows is multiplied by S as in the following equation where $b=60$ and constant $f$ can vary from 0 to 36:

$$R = \begin{pmatrix} w_{i,0} & \cdots & w_{i,f-1} \end{pmatrix} \begin{pmatrix} s_{0,0} & \cdots & s_{0,b-1} \\ \vdots & \ddots & \vdots \\ s_{f-1,0} & \cdots & s_{f-1,b-1} \end{pmatrix}.$$

The original implementation of beam forming had three inefficiencies with respect to SIMD parallelism:

1. It performed more FLOPS than necessary
2. It did not combine multiplications and addition into fused multiply-add (FMA) operations.
3. It performed many permutation operations, while a Haswell processor has only one execution unit for processing these.

The improved algorithm has corrected all three and has been implemented in AVX2 intrinsics. It provided speedups of up to x1.5 over the previous implementation. Figure 5 illustrates the beam forming processing pipeline (redrawn from textbook [7]).

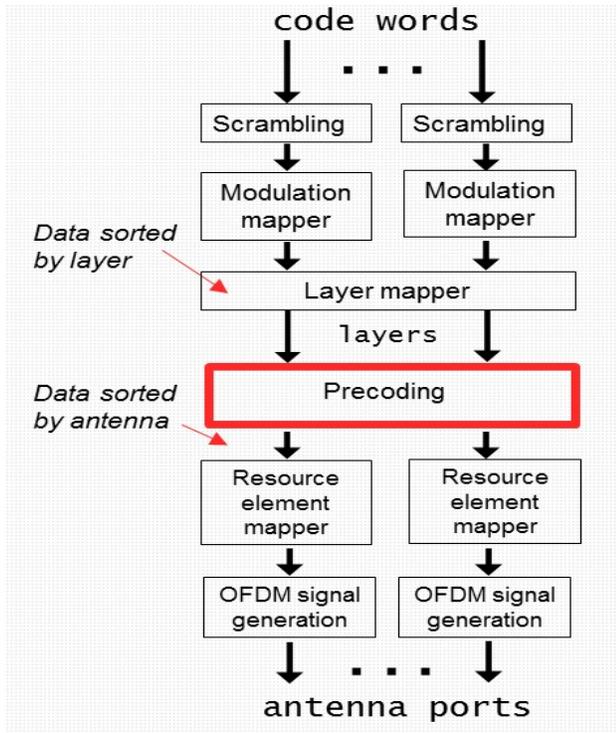

Fig. 5. Beam forming

Turbo decoding has also been re-implemented, tested and given acceleration factors of 1.7 to 1.9 for its pipeline stage. The corresponding number of CPUs required for a parallel multi-stream execution of multiple instances of turbo decode has been reduced from 20 to 12, with similar energy savings.

IV. IMPROVING PROGRAMMER PRODUCTIVITY: A GENERIC TOOL

The 5Gperf project has also designed and implemented a new software tool called *the optimizer* so that programmers can convert performance-naive MATLAB code to optimized C in much less time and in a reliable fashion. It provides high programmer productivity, highest-possible performance for the predefined operations it applies and portability to ARM architectures through Numscale's bSIMD library [4, 5]. The optimizer's principle and design is summarized by figure 6. The application developer in charge of the signal-processing pipeline's algorithms can transform a high-level algorithm description to optimized and portable C code by annotating critical portions of his code. Each such code segment corresponds to an algorithm building block available in the optimizer's database. The optimizer tool then replaces the annotations by the most efficient version of the building block on a given matrix size and target architecture.

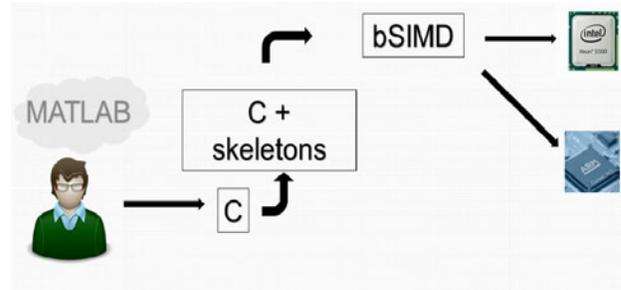

Fig. 6. The 5Gperf software tool

The optimizer's parser looks for pragma instructions in the source C code.

   /// PRAGMA INCLUDES

This line tells the optimizer to put extra includes required by building blocks where it appears.

   /// PRAGMA FUNCTIONS

This line tells the optimizer to put extra codes required by building blocks where it appears.

   /// PRAGMA BEGIN ke, p2, p3, . . . , pn

This line tells the optimizer to insert optimized code produced by the kernel named ke.

The following example is valid input for optimizer.py:

```
1.   #include <iostream>
2.   #include <cstdio>
3.   /// PRAGMA INCLUDES
4.   /// PRAGMA FUNCTIONS
5.   int main() {
6.   std::cout << "begin" << std::endl;
7.   /// PRAGMA BEGIN algo, b, c,
8.   std::cout << "BAD" << std::endl;
9.   /// g, h
10.  std::cout << "BAD" << std::endl;
11.  /// PRAGMA END
12.  std::cout << "end" << std::endl;
13.  return 0;
14.  }
```

On the above example the optimizer will replace lines 6-10 by the best algorithm given by the executable bbs/algo with parameters b, c, g and h. The choice of best algorithm depends on vector sizes for SIMD libraries, loops and target architecture. The parameters are passed to the bbs/algo executable as command line arguments.

The output will look like this:

```
1.  #include <iostream>
2.  #include <cstdio>
3.
4.  #ifdef OPTIMIZER_ACTIVATED
5.    // Extra includes here required by building blocks
6.    // This block of code replaces the original ///
       PRAGMA INCLUDES
7.  #endif
8.
9.  #ifdef OPTIMIZER_ACTIVATED
10.   // Extra code here required by building blocks
11.   // This block of code replaces the original ///
      PRAGMA FUNCTIONS
12. #endif
13.
14. int main() {
15.   std::cout << "begin" << std::endl;
16. #ifndef OPTIMIZER_ACTIVATED
17.   std::cout << "BAD" << std::endl;
18.   std::cout << "BAD" << std::endl;
19. #else
20.   // OPTIMIZED CODE HERE PROPERLY INDENTED
21. #endif // PRAGMA BEGIN line 6
22.   std::cout << "end" << std::endl;
23.   return 0;
24. }
```

## V. Conclusions

Even highly optimized signal-processing code can be improved by moderate factors for such a critical application as 5G signal-processing. But the process is very work intensive, especially because the compute tasks are small and memory access is a limiting factor. So a special type of generic programming is needed. Programmer productivity should be multiplied and useful speedups (20% to 100% accelerations) can be obtained on all the signal-processing pipeline.

## Acknowledgment

Guillaume Quintin and Sylvain Jubertie of Numscale were the main developers of the optimizer software tool and contributed to algorithms. The authors thank Yang Ganghua and Bill McColl for initiating and supporting the 5Gperf project. Antoine Petitet, Alain Dominguez and Chong Li were involved in some of the technical decisions and made suggestions that contributed to the project's success.